# Turning Time from Enemy into an Ally Using the Pomodoro Technique

Xiaofeng Wang[1], Federico Gobbo[2] and Michael Lane[3]

[1] Lero, The Irish Software Engineering Research Centre
Limerick, Ireland
xiaofeng.wang@lero.ie

[2] DICOM DIpartimento di Informatica e Comunicazione, University of Insubria
Via Mazzini 5 21100, Varese, Italy
federico.gobbo@uninsubria.it

[3] Computer Science and Information Systems, Univeristy of Limerick
Limerick, Ireland
michael.lane@ul.ie

**Abstract.** Time is one of the most important factors dominating agile software development processes in distributed settings. Effective time management helps agile teams to plan and monitor the work to be performed, and create and maintain a fast yet sustainable pace. The Pomodoro Technique is one promising time management technique. Its application and adaptation in Sourcesense Milan Team surfaced various benefits, challenges and implications for distributed agile software development. Lessons learnt from the experiences of Sourcesense Milan Team can be useful for other distributed agile teams to turn time from enemy into an ally.

## 1. Introduction

Time is a priceless and scarce resource for software development projects [4]. It is especially true in agile software development. A brief review of the 12 agile principles behind the Agile Manifesto reveals that time is an important dimension of agile processes, symbolized by terms such as "early", "frequently", "couple of weeks", "daily", "regular intervals" in these principles [1]. Agile teams work with short time-boxed iterations and need to maintain a fast yet sustainable pace throughout the project lifespan [3]. When moving to a distributed setting, the time dimension is further complicated by issues such as time zones, geographical distance, and different cultures [2]. However, there is very little reported evidence of effective time management techniques applied in agile software development, especially in the context of distributed teams.

The Pomodoro Technique is a time management tool that was originally intended to optimize personal work and study. More recently, it has been widely applied by Italian agile teams [9]. Awareness of this technique is growing among the wider, international agile community (two tutorials on the Pomodoro Technique have been given in Agile 2009 - the international conference). The technique is named after the usage of a common kitchen timer in the shape of a tomato ("pomodoro" in Italian, see Figure 1). The heart of the Pomodoro Technique is 25 minutes of focused, uninterrupted work on one task, then 5 minutes of rest. There are also rules to keep the integrity of pomodoro, and tactics to deal with internal and external interruptions. However, starting as a personal time management tool, how is it applied by an agile team, especially when the team is working in a distributed environment? There is no ready answer in spite of the increasing popularity of the Pomodoro Technique in the agile community.

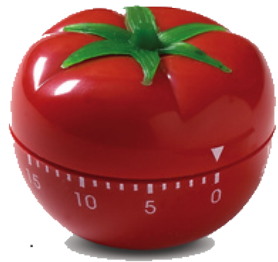

**Figure 1:** A tomato-shaped timer

Based on this observation, the objective of our study is to provide a better understanding of the application of the Pomodoro Technique in agile teams, especially when they work in distributed contexts. To this end, we studied in-depth one agile team that has applied the Pomodoro Technique extensively. The team collaborates with other remote sites of the company where the Pomodoro Technique is not used. This allows us to reflect on the impact of the Pomodoro Technique (and the lack of it) in a distributed context.

The remaining part of the chapter is organized as follows. In the next section we review a set of time-related issues and argue the importance of time management in software development in general and agile software development in a distributed context in particular. It is followed by an introduction of the Pomodoro Technique. Then the experience of Sourcesense Milan Team using the Pomodoro Technique is presented. We analyse their experience and provide useful guidelines for implementing the Pomodoro Technique in the following section. The chapter ends with a conclusion section that highlights the contribution of our study and points out future studies.

## 2. Time is An Enemy?

Time occupies a crucial place in software development projects. It is one of the most important factors dominating software development processes [10]. And just like many people have experienced, to various degrees, the anxiety associated with the passage of time, many software development projects have suffered from a set of time-related issues. Back in 1975, Brooks claimed that "more software projects have gone awry for lack of calendar time than for all other causes combined" [4]. In [10] a set of interlinked time-related problems in software development processes is summarised, including *bottlenecks*, which occur when one or more functions in the development process are dependent upon the output of another function within the process, resulting in developers having nothing to work on in the meantime; *schedule problems*, both construction of a feasible project schedule and to meet the schedule that has been set; *difficulty in time estimation* of large module/class/task; *time pressure*, which happens typically towards the end of a development process when the development team cannot meet the project schedule either because of poor time estimations or bottlenecks; and *late delivery*, which occurs as a result of inappropriate project planning, usually due to poor estimations.

These inter-related problems, in essence, are all subjects of time management, one of the major knowledge areas of project management [13]. This area involves decomposition of project work into manageable tasks, estimation of task durations, scheduling of tasks, and controlling and monitoring the execution of tasks. Effective time management is crucial for addressing the time-related problems and leading to the success of software development projects. However, [10] argue that in early software engineering projects, when waterfall versions started to emerge, time issues were essentially neglected. As we proceed along the software engineering timeline, time issues receive increasingly more attention and their importance in software development processes is increasingly recognized, as evidenced in Team Software Process (TSP) [12], Rapid Application Development (RAD), and recently in agile software development.

Time plays a more crucial role in agile software development than in conventional waterfall-like software development processes. This is demonstrated by a review of the 12 agile principles [1] from a time perspective. A review of these principles shows that 50% of them have an emphasis on time:

- Our highest priority is to satisfy the customer through **early** and continuous delivery of valuable software.
- Welcome changing requirements, even **late** in development. Agile processes harness change for the customer's competitive advantage.
- Deliver working software **frequently**, from **a couple of weeks** to **a couple of months**, with a preference to the **shorter timescale**.
- Business people and developers must work together **daily** throughout the project.
- Agile processes promote sustainable development. The sponsors, developers, and users should be able to maintain **a constant pace** indefinitely.
- **At regular intervals**, the team reflects on how to become more effective, then tunes and adjusts its behavior accordingly.

These principles pose new time-related challenges to agile teams, apart from the previously discussed time issues. Agile teams need to work at a fast yet sustainable pace. The risk in having sole focus on velocity of development is the reduction of enthusiasm among team members. This may then have an impact on the sustainability of an agile team's daily work. It is important to achieve this equilibrium in agile software development, but it is underestimated in practice [9]. As a consequence, time management in agile software development not only means effective planning and monitoring of the work to be performed, but also should help agile teams to create and maintain a fast yet sustainable pace.

These two aspects of time management in agile software development can be further complicated as software development moves into global, distributed settings, which is an approach adopted by many current software projects. Issues such as time zones [6], geographical distance [7], and different cultures [14] will affect the previously discussed time-related issues and the effectiveness of time management techniques employed by agile teams. To better understand the impacts of distributed context, it is useful to consider different distributed team configurations (see [8] for different kinds of team configurations). Each different team structure presents different benefits to the work being undertaken, and would impact the time management technique used by agile teams in different ways.

In spite of the importance of time and time management in agile software development, however, there is a paucity of both time management techniques and studies on the application of these techniques in agile software development, let alone in a distributed agile context. The Pomodoro Technique is one promising time management technique and is increasingly popular in the agile community. A growing number of agile teams use the pomodoro technique within their agile development processes [9].

## 3   The Pomodoro Technique

The goal of the Pomodoro Technique is to encourage consciousness, concentration, and clarity of thought through effective time management. A 'pomodoro' is 25 minutes of focused, uninterrupted work on one task then 5 minutes of complete rest. The inventor claims that, based on scientific proof, "20- to 45-minute time intervals can maximize our attention and mental activity, if followed by a short break" [5]. The 5-minute break aims to support team members in establishing and maintaining an optimal attention curve while engaged in project activities. In order to increase the impact of this effect, following every four consecutive pomodoros a longer pause of 15 minutes is recommended.

The technique can improve the productivity of an individual. Improvements in productivity are achieved through increased motivation and the technique has also proved effective in supporting the management of complex situations. These benefits can be achieved through the following two inter-related aspects of the Pomodoro Technique: *time-boxing* and *duration estimations*.

### 3.1  Pomodoro as Time-box

One of the primary inspirations behind the Pomodoro Technique is time-boxing [5]. Time-boxing suggests that, once a series of activities has been assigned to a given time interval, the delivery date for these activities should never change. If necessary, the unfinished activities can be reassigned to the following time interval. Corresponding to the idea of time-boxing, to maximize participant concentration, every time-box (or pomodoro), needs to be protected. "Protecting pomodoro" leads to fewer interruptions. There are two kinds of interruptions that need to be addressed:

- *Internal*: These interruptions are triggered by the participant, e.g. "I should check email", or "I need to get a coffee";
- *External*: Triggered by other entities, e.g. a phone call or a request from a colleague.

In order to handle these interruptions effectively, an "indivisible rule" needs to be enacted. A pomodoro represents twenty-five minutes of pure work that cannot be split up. There is no such a thing as a half or a quarter of a pomodoro. If a Pomodoro is interrupted definitively, i.e. the interruption is not deflected, then the pomodoro is considered to have never commenced – it is made void.

Used as a time-boxing tool, the Pomodoro Technique can help enhance focus and concentration on work by cutting down on interruptions. Consequently, it can alleviate anxiety linked to the passage of time and reduce both time-wasting and overtime. A sustainable working pace can be obtained through the alternation of work and rest and the combination of short breaks and long pauses.

### 3.2 Pomodoro as Unit of Effort

To master and improve the use of the Pomodoro Technique, an underlying daily process is suggested, which consists of five stages:
- *Planning*: to decide the activities to do in the day;
- *Tracking*: to gather raw data on the effort expended and other metrics of interest;
- *Recording*: to compile an archive of daily observations;
- *Processing*: to transform raw data into information; and
- *Visualizing*: to present the information in a format that facilitates understanding and clarifies paths to improvement.

In each stage, the pomodoro plays the role of unit of estimation. Two rules apply: 1) if a task lasts more than 5-7 pomodoros, break it down. Complex activities should be divided into several activities; and 2) if it lasts less than one pomodoro, add it up. Simple tasks can be combined.

Used as an effort estimation tool, the Pomodoro Technique can support the refinement of an effort estimation process through the use of continuous reflection of team activities (more detailed instructions of how to apply the Pomodoro Technique as a personal time management technique can be found in [5].

The Pomodoro Technique was invented initially for individual work, as a personal time management tool. But it has been developed and refined in the context of teamwork by the inventor and advocates of the technique over the time. [9] report the technique as a *team* time management tool used by several XP teams. They claim that the Pomodoro Technique is an unstressful - as well as efficient - way to help teams find their "natural" rhythm in daily work. Their study is a good starting point and provides a broad picture for investigating how the Pomodoro Technique can be applied in agile teams. Our study intends to go into more depth to understand the benefits, issues and concerns of using the Pomodoro Technique in an agile team within a distributed context.

## 4 The Application of the Pomodoro Technique in Sourcesense Milan Team

In this section, we present one case study of the application of the Pomodoro Technique in an agile software development team. More specifically, we investigate how Sourcesense Milan Team, an XP team, applied the Pomodoro Technique as both a time-boxing tool and an estimation tool. We also present their reactions to working in a distributed setting, collaborating with other locations that did not use the Pomodoro Technique.

### 4.1 Background of Sourcesense Milan Team

The Company, Sourcesense, is a European systems integrator providing consultancy, support and services around key Open Source technologies (Table 1).

**Table 1: Company Overview**

| Company: Sourcesense | |
|---|---|
| **Number of developers** | < 50 |
| **When was agile introduced** | 2007 |
| **Domain** | Open Source Systems integrator |

Sourcesense Milan team is a part of the company that is distributed across several countries, as illustrated in Figure 2.

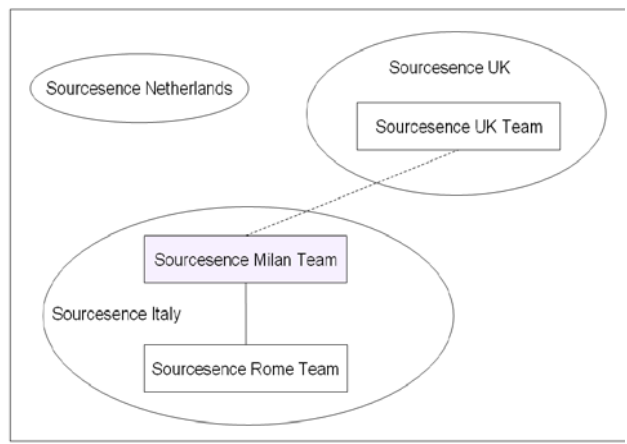

**Figure 2:** The distribution context of Sourcesense Milan Team

In Italy, there are two teams that are focused primarily on software development: Sourcesense Milan Team and Sourcesense Rome Team. Sourcesense UK contains a team that plays the role of customer proxy: Sourcesense UK Team. The remaining part is based in the Netherlands, Sourcesense Netherlands, and it is responsible for the provision of consultant-oriented services. In the context of this case study, Sourcesense Netherlands is not engaged in any of the projects performed. Table 2 is an overview of the distribution context that Sourcesense Milan Team is embedded in. In terms of configurations of distributed teams, in the context of this case study, two structures are in place:

- *Modular*: Sourcesense Milan Team and Sourcesense Rome Team take responsibility for separate modules or features of the system under development.
- *Functional expertise*: Sourcesense UK Team is deemed to be the most appropriate source for requirements elicitation in the absence of customer access.

**Table 2: Team Overview**

| Locations | Number of members | Roles |
|---|---|---|
| Sourcesense Milan | 10 | Developers, XP coach |
| Sourcesense Rome | 3 | Developers, project manager |
| Sourcesense UK | | Customer proxy |

Sourcesense Milan Team, formed in February 2007, is composed of experienced developers, well known in the Italian Agile community. The agile methodology the team uses is XP. Some of the members joined Sourcesense after working in other agile teams since 2000, and were trained by the inventor of the Pomodoro Technique. There are 10 developers in the team, among them one playing the role of XP coach. Four of the developers are part-time. The development team is concurrently working on multiple projects, which have different customers, operate in different domains and leverage different technologies. Multi-tasking in multiple projects renders estimation and time allocation among projects more crucial and challenging.

## 4.2 The Development Process of Sourcesense Milan Team

Sourcesense Milan Team works at a fast pace, restricting itself to 1-week iterations. An iteration planning meeting is held where the work in the current iteration is planned out. User stories are selected for implementation in the iteration and the relevant story cards are posted on the white board in the team's office. The team also uses online spreadsheets to track the progress of user stories. Because the team does not have an on-site customer, a progress report to the customer detailing all achievements/issues of the previous iteration is compiled and sent out to the customers after the iteration planning meeting.

Generally every two weeks the team conducts a retrospective where members reflect on their development process. Each team member is given an opportunity to lead retrospectives, not just the coach.

Every Wednesday is set aside for research activities – the team use their time to focus on the study of both project-related concepts and growth of general work-related skills.

A standup meeting is conducted at 9:30 everyday, generally lasting for no longer than 15 minutes. The first activity performed by team members is to individually review the previous day's journal. This is a document produced by team members at the end of each day outlining the activities performed during the day. It serves multiple purposes: providing non-collocated team members an immediate view of overall team progress, enabling team members to achieve closure on their day's work and acting as a review that is used to set the tone for the following day's work. Journal reviews are completed before 9:30 each morning as this is the time scheduled for the daily standup meeting. Having used the journal to review what was achieved on the previous day, the two main questions addressed in the standup meeting are: 1) "What am I going to do today?" and 2) "What are the problems in doing it?"

After the standup meeting, the team goes into development mode. The developers work in pairs all the time. Pair rotation happens regularly.

The development process of Sourcesense Milan Team would have been similar to many other XP or agile teams but for their use of the Pomodoro Technique. This is seen more clearly by reviewing the perspectives of Sourcesense Milan Team in relation to the impact of their use of the Pomodoro Technique. Two main areas are presented:
- Having the pomodoro play two key roles in the development approach: "timebox" and "unit of effort".
- Issues arising from collaborations with distributed teams that do not use the Pomodoro Technique.

### 4.3 Pomodoro as Time-box

Pomodoro is used to time-box the development activities of the team. Several aspects of using pomodoro as a time-boxing tool in the team are highlighted below.

#### 4.3.1 One Pomodoro Rules Them All

The initial application of the Pomodoro Technique involved a pomodoro timer for every pair of developers. The owner of the card was responsible for loading the timer, and updating the card. As a result of the responsibility assumed by each partner within a pair to support their colleague, internal interruptions were minimized. The coach was responsible for deflecting external interruptions – he protected the integrity of each pomodoro as the pair worked on its activities.

As a result of retrospectives on how best to effectively apply the Pomodoro Technique in a team setting, it was decided to experiment with an extension of the Pomodoro Technique. The name given to the extended approach was "shared pomodoro". This approach proposed that a pomodoro be applied to the whole team, i.e., "one pomodoro rules them all" (Figure 3 is an illustration of a team working with a shared pomodoro timer).

Although some developers expressed reservations on this approach initially, it is now accepted by the team as a good way to address certain issues that were having an adverse impact on the team. These issues were related to interruptions that were being caused between pairs of developers. When each pair had their own pomodoro and their own associated pomodoro timer, there were cases where they disturbed each other due to different break times. Some pairs preferred to work for 50 minutes and take a longer pause thereafter; others used pomodoro just to track their work without paying attention to the breaks. This resulted in noise and distractions for pairs that were not at break. The shared pomodoro approach ensures that when a pomodoro is being tackled, everybody is working; no one has distractions from other people having a pause. It is also good for the whole team to have long pauses all together. Synching up the pomodoros can also increase cross-pair communication after breaks. Alignment of breaks enables the team to switch pairs more frequently. When pomodoros were not aligned, it was not possible to switch partners within pairs and there was an increased risk of a pair attempting too much in one session. The shared pomodoro approach also helps the team members to be more disciplined in their adherence to both working time and break time. This helps the team to obtain a working rhythm. The team actually behaves like a team, and the cohesion between the team members is increased. An additional benefit to the shared pomodoro approach is that having breaks shared with other pairs acts as a motivational device. This is especially true for pairs that find themselves less than enthusiastic about returning to work due to the types of tasks that are required to be done in their pomodoros at a given time. Not everybody can get to work on the "cool" tasks all the time – but having the opportunity to share in the overall environment at break time helps to overcome this issue.

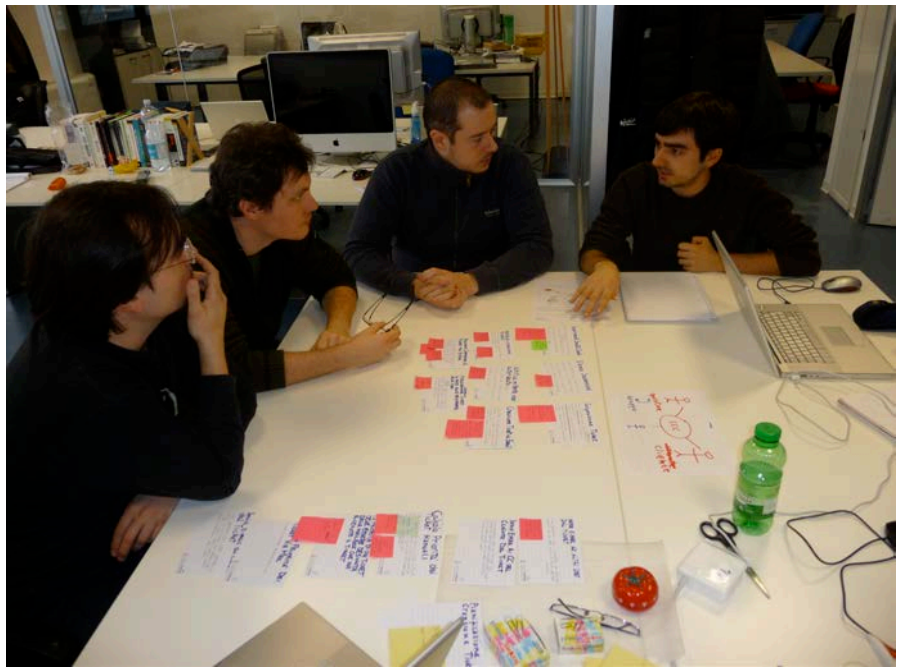

**Figure 3:** Team engaged in Planning using the shared pomodoro approach

The team also realizes that shared pomodoro should not be used to block potentially constructive interactions during a 25-minute time slot. They believe that the Pomodoro Technique and shared pomodoro in particular, is intended to make the team members more aware and respectful of everyone's work. Interruptions that should be avoided include phone calls, instant messages, and people wandering in and requesting assistance from team members on non-project-related activities. However, as stated above, there are acceptable interruptions. The team works in a face-to-face setting in an open office. The coach observed that if he hears someone having trouble he would intervene and help them. He believes that it might be worthwhile to interrupt his work for a minute to save the other pair from maybe 30 minutes of "*puzzling over something that I know straight away*".

### 4.3.2 Break is Break

The team realizes that the 5-minute break is as important as the 25-minute working time for them. The team is well aware of the fact that breaks may be used in a variety of ways. The uses of a 5 minute break can be viewed as a continuum, ranging from complete relaxation (day dreaming, practicing Qigong, sleeping), to more active breaks, such as responding to emails, reading blogs, or even discussing what is just been completed in the last pomodoro with colleagues. According to the team, it takes training to rest effectively, just like it takes training to be able to develop software. In accordance with the recommendations of the Pomodoro Technique the team treats the breaks seriously and the developers are not encouraged to do activities during the break. For example, the team keeps a personal machine separate from work machines. This computer is situated in a different location than the work machines. The intention is to discourage team members from checking emails and reading blogs during a pomodoro. However, it should be noted that team members do check emails from time to time on their laptops. The coach admits that, even though it is a sign of indiscipline, it does happen occasionally.

### 4.3.3 Time-boxing Non-development Activities

Pomodoro is also used by the team to time-box not only development, but also other activities, such as study and meetings.

As previously stated, Wednesday is scheduled for team study. The team spends four pomodoros studying various topics on Wednesday afternoons. To the team, study is not a break – customer-focus is maintained during these activities. One of the goals of study activities is to promote lateral thinking - "*to try to solve the problem from different angles*". The team needs to get quick feedback on what is learnt. Therefore, stories used to drive study activities should be as small as possible in order to shorten feedback cycles. Time-boxing study time with the Pomodoro Technique could help. Another application of this technique is meetings. These events are also time-boxed with pomodoro, in order to help people focus and reduce wasted time.

## 4.4 Pomodoro as a Unit of Effort

Pomodoro is used extensively, as a unit of effort, in planning, estimating and tracking progress. In the case of Sourcesense Milan Team, the unit of effort is one pomodoro per pair. For example, there are 6 full-time people in the team (3 pairs). If each pair works on 10 pomodoros per day, the total team capacity is 30 pomodoros per day.

Activities that do not require team members to work in pairs, such as administrative meetings, are measured by half-pomodoro per person. For example, if a meeting takes 5 developers half an hour, the total effort the team spends on the meeting counts for 2.5 pomodoros.

### 4.4.1 Pomodoros vs. Abstract Story Points

The team originally estimated user stories in abstract story points, which is a common practice suggested by XP [3]. A story point is an abstract complexity measure of a user story. Initially, the team did not commit to the estimation of durations for each story point. In time, the team switched to directly estimating everything in pomodoro per pair. To the team, story points started losing meaning. Instead, the pomodoro became a more concrete measure of effort.

Another concept utilized by the team, following the suggestion of the original Pomodoro Technique inventor, was "pomodoro type". This concept classifies pomodoros spent into different categories, such as analyzing, coding, refactoring and testing, to help the team to obtain a better understanding of the effort spent on different types of work, e.g., the ratio of coding over refactoring, or how much effort spent on "*wrestling with the server*" in deployment.

### 4.4.2 Tracking Pomodoros

Every morning the team members pick up the story cards from the whiteboard (as shown in Figure 4) and the current story cards that are on the desk of the developers. Usually each card holder marks a cross on the back of the story card for each pomodoro consumed. This helps to track how much real effort was spent on a story. Every evening they will put all the cards together again on the whiteboard, to understand how many pomodoros of user stories are left unfinished, and if they can finish them within the iteration. It also helps to measure how much uninterrupted work the team can do in a day. They have a shared spreadsheet to store historical data about each iteration. The white board only shows the current iteration.

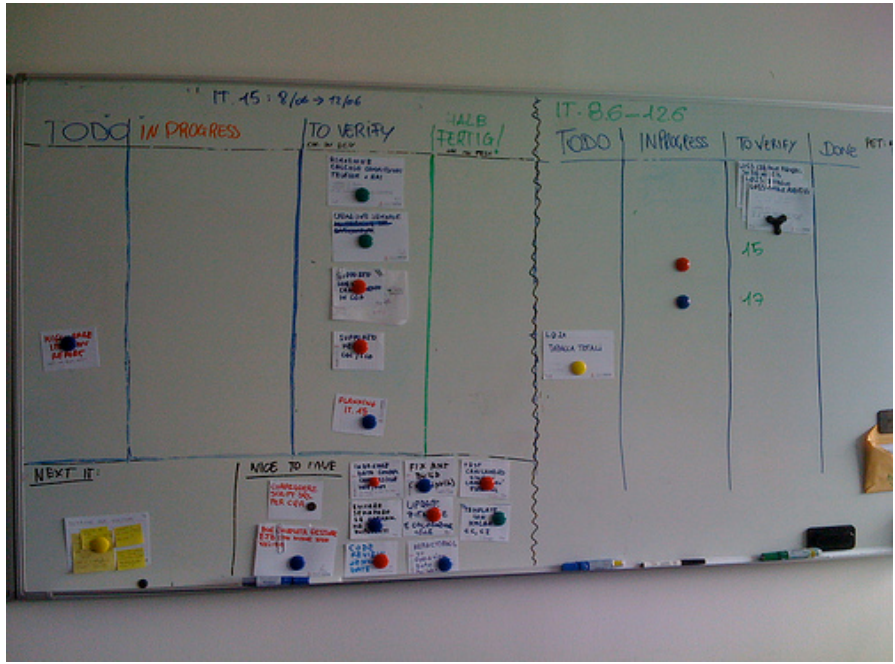

**Figure 4:** The whiteboard with user stories in current iteration

### 4.4.3 When Not to Measure or Rrack with Pomodoro

The team realizes that certain activities do not require estimation and also may not require that the effort spent on them be tracked. The typical activity that the team does not track is non project related exploration. There are two types of exploration: project related, called spikes and non project related, more exploratory study that are driven by the interests of the team members rather than by project issues. Spikes, which are related to any user story, are tracked using the Pomodoro Technique. In contrast, the team believes that general explorations, which are described as "*useful exercises of dreaming where you would like to be or to do*", should be conducted without time pressure. The team often conducts experiments collectively on the code base without the objective of production work. They believe that the result of the exploration should be shared among the team members. Therefore, presentations of new technology and new ideas are held from time to time. There is also the concept of a "lunch cinema club". Team members get together and watch videos from important agile authors on the big TV. When time permits, these viewing sessions are then followed by a short discussion. All these activities are not tracked in pomodoros.

### 4.5 Addressing Remote Collaboration With Teams That Do Not Employ The Pomodoro Technique

Sourcesense Milan Team operates in a larger, distributed environment, as previously described. However, the Pomodoro Technique has not been adopted by the other sites. Sourcesense Milan Team members feel that the use of the Pomodoro Technique to pace and time-box each day "*protects*" them from interruptions, because they believe that "*two or three chats open, reading emails while working - you can't say it's concentrated work*". The application of the Pomodoro Technique also enhances the teamness of Sourcesense Milan Team. The team members feel that they are working as a real team, where internal interruptions are minimized and the coach can protect the team from external interruptions. They also feel that their application of the Pomodoro Technique makes the project work more transparent. They are able to demonstrate to other sites how many hours of the day are spent on development, which helps to build the trust of other sites on Sourcesense Milan Team.

The application of the Pomodoro Technique in Sourcesense Milan Team created a team working style which could be different from the other sites where the Pomodoro Technique is not used, which in turn may enlarge the mismatch of team cultures at different sites. For example, although Sourcesense Rome Team is located within the same national borders as Sourcesense Milan Team and functions as a development site as well, the site is felt to be "*just as distant as the Netherlands*". The fact that the Rome Team does not apply the Pomodoro Technique may have an adverse influence on this perceived distance..

The other distributed sites (Netherlands and UK) are more engaged in consulting activities rather than software development. The nature of the work decides that many of the staff there works as individuals rather than in teams. In such a context, it is more difficult for them to be free from all sorts of interruptions. They are not as "protected" as Sourcesense Milan Team by the Pomodoro Technique.

However, Sourcesense Milan Team is aware of the existence of different working styles and team cultures at the different sites, and believes that they should not impose the Pomodoro Technique on the other sites. They are concerned that their use of the Pomodoro Technique may act as a barrier to effective collaboration:

> "*The difficulty is in communication, being reliable, and trying to make yourself useful. You have the risk of presenting a wall like 'this is our method, we do it in this way, and you have to work around us'. This is not what the method is meant to be. So you have to learn to be humble enough to be helpful to your colleagues, not just with your customers*".

## 5   Turning Time into an Ally

In this section we reflect on the experiences of Sourcesense Milan Team using the Pomodoro Technique, and draw some practical implications.

## 5.1 Shared Pomodoro

The shared pomodoro concept used in Sourcesense Milan Team is an adaptation of the Pomodoro Technique that was invented as a personal time management tool originally. The team has obtained positive results using shared pomodoro. However, although not specifically highlighted in the case study, it should be noted that there are several potential drawbacks to the use of this technique. The working rhythm of the team is liable to be interrupted by any team member's necessary and unavoidable pause during a pomodoro. It takes more effort to coordinate the whole team to start the first pomodoro when people come in to the office at different times in the morning. Similarly, commencing a new pomodoro after short or long breaks may be delayed. The whole team has to wait for all the team members coming back to the open space. It is possible that one pair may be working on a really challenging issue, and following completion of the pomodoro the may need a slightly longer break than others. Conversely, another pair may work on a relatively easy task and wish to take a shorter break. These different physical and psychological impacts are not easily reconciled by shared pomodoro.

*Practical Tip*: To understand whether shared pomodoro is right for a team, the team needs to know how confident each individual team member is with the rhythm set by using the basic Pomodoro Technique. If they are already using it in a disciplined manner and are comfortable in the environment, then the team can try shared pomodoro for 4 – 6 weeks. At that point, they may then decide whether or not to adopt this extension of the technique. Consultation with the team about adoption of this approach is preferable to managerial imposition of it as a mandatory process. Otherwise, there is a risk of making the developers uncooperative, because they could feel their freedom and creativity is being constrained by a shared timer.

The experience of Sourcesense Milan Team also reveals that the Pomodoro Technique should be used to block internal or external interruptions, not constructive interactions among team members. Communication and interaction are a key tenet of any agile method. The Pomodoro Technique helps a team to be aware of positive interactions and supports the deflection of unnecessary and unconstructive interactions/interruptions. However, it is up to the team to use the technique in a sensible way to balance maximum concentration of the team and effective interactions among the team members.

## 5.2 Collective Breaks

Breaks are taken seriously by Sourcesense Milan Team and the team members are encouraged to relax and recuperate during their break. Using shared pomodoro can help the team members to take breaks regularly, but does not guarantee that team members take proper breaks. The very fact that the team members are taking breaks altogether would tend to make breaks more like the extension of a pomodoro. A simple chat could easily slip into a technical discussion of the work just completed in the previous pomodoro. Such a situation could result in the break being just as taxing as the work itself. The case of Sourcesense Milan Team presents some good practices to foster relaxing breaks, such as placing the personal machine far from the working machine.

*Practical Tip*: In Sourcesense Milan Team, the team members also take breaks individually. This does not specifically promote relaxation. The shared pomodoro concept can be extended to breaks as well. A collective break can be taken where team members do some relaxing activity together during breaks, such as game playing, 5-minute fitness club, etc. Collective breaks can help to avoid the situation where a pair of developers does not wish to let go of their work following the signal from the pomodoro timer.

### 5.3 Estimation and Tracking

Sourcesense Milan Team uses a pomodoro per pair as the unit of effort in estimation and tracking. Pomodoro is actual time measure in contrast to abstract story points, or "gummy bears". Consider the purpose of estimation and tracking in agile software development. Estimation of user stories is more about granularity of tasks than about effort needed to implement them, and tracking effort spent on them is less about amount of work done than about learning and improving the team's capability of estimating throughout. Using actual time makes estimation and tracking more transparent and learning more concrete.

However, estimates and tracking results should not be used as measurement of each individual team member's performance, otherwise there will be a tendency to game the number, and the Pomodoro Technique used as a Tayloristic approach to exploit the team or a micro management tool for project management.

### 5.4 One Pomodoro Rules All Sites?

In Sourcesense Milan Team, the Pomodoro Technique is not used in a truly distributed manner. But the technique is not confined to co-located teams and can be adapted and used in distributed settings and in a distributed manner. Just as the expansion of the Pomodoro Technique from a personal time management tool to a team management tool surfaced various complexities, various benefits and challenges are likely to emerge when the technique is applied in a distributed context.

*Practical Tip*: Remote pairs or distributed team members could apply the shared pomodoro approach as a time-pacing tool to synchronize and coordinate their activities. It can help better manage interruptions that are generally an indispensible element of distributed development, such as emails, instant messages and phone calls. The shared pomodoro approach can be implemented in distributed teams using virtual space and virtual timer technologies (a virtual timer on a server), whereas pauses can be shared via Skype or analogue systems.

If distributed team members keep their individual pomodoro timers, it would be more difficult for them to be aware if others are in the middle of a pomodoro or not, since the working status is not as obvious as in a co-located team where it can be understood at a glance. To address this issue, the working status needs to be made more explicit and visible to each of the distributed members. For example, cherrytomato (http://www.chrylers.com/cherrytomato/), a software tool that intends to support the distributed pomodoro technique, integrates a virtual pomodoro timer with instant messaging tools such as Skype. When a developer starts a new pomodoro, his status in Skype would be switched to "do not disturb" automatically and can inform other team members how many minutes are left for the current pomodoro or show other customized messages.

The experiences of Sourcesense Milan Team also indicate that, if used properly, the Pomodoro Technique can increase the transparency of both estimates and project-related work expended at different sites. Consequently, this may help to build up trust among different sites. There are a growing number of agile planning tools that support distributed estimation and tracking. The Pomodoro Technique can be easily integrated in these tools to support estimating and tracking in pomodoros.

However, the issues that are associated with shared pomodoro in a co-located team may become more challenging when temporal, geographical and social/culture distances are involved. For example, it will take greater effort to coordinate the whole distributed team to start the first shared pomodoro. In instances where the team is distributed across different time zones, developers will be involved in different phases of a working day, resulting in possible variances in concentration levels between team members. Therefore, the balance of maximum concentration of team members and effective interactions among them may be more difficult to maintain in a distributed context. Last but not least it needs to be cautioned that the Pomodoro Technique should not be used as a Tayloristic approach to exploit outsourced sites.

## 6  Conclusions

In this chapter we presented a case study of the application of the Pomodoro Technique in an agile software development team. The effects of the technique on the team were analysed through two angles: pomodoro as timebox and pomodoro as unit of effort. We argued the benefits and potential issues of using one pomodoro for the whole team, team breaks, estimating with pomodoro (real time) rather than abstract points. We also explored the scenarios where the Pomodoro Technique should not be applied. If and how the technique can be applied in a distributed setting was also examined. Even though the Pomodoro technique is not used in a truly distributed manner in the case we studied (which is a major limitation of our study), we believe that the technique itself is a welcome addition to the agile development toolkit.  We would hope that the lessons learnt from this case study could be easily adapted into distributed settings.

Our study contributes to the body of knowledge of an under-developed theme within agile research: effective time management in fast-paced agile software development. The practical implication of our study is a better understanding of time management in agile software development, and several concrete suggestions of how to effectively apply the Pomodoro Technique and make the best out of it in an agile team working in a distributed setting. Our study is just the first step in the investigation of interesting phenomena arising from the application of the Pomodoro Technique in agile teams. Our report of the case is presented from the perspective of the agile team that used the Pomodoro Technique, and claimed it to be an effective approach. To further establish the effectiveness of this technique, more objective assessment is needed, and the viewpoints of all stakeholders related to the team need to be obtained. During this research, other issues emerged as candidates for future exploration., including how to encourage team members to take a break, how to use timers properly in an open office (visible vs. invisible timer, sound of ticking and ringing, and mechanical timer vs. digital timer vs. software tool), interesting secondary effects associated with breaks, and learning associated with using pomodoro for estimation.

**Acknowledgements:** Special thanks go to Matteo Vaccari and his Sourcesense Milan Team who collaborated on our study and supported the production of this book chapter. Their experiences with the Pomodoro Technique were the inspiration of our study.

## References

1. Agile Manifesto: http://www.agilemanifesto.org/, 2001, last visit Nov. 2009.
2. Ågerfalk, P.J., Fitzgerald, B., Holmstom, H., Brian Lings, B., Lundell, B. and , and Ó. Conchúir, E.: A Framework For Considering Opportunities And Threats In Distributed Software Development. Proceedings of the International Workshop on Distributed Software Development (DiSD 2005), Paris, 29 August 2005: Austrian Computer Society, 47–61.
3. Beck, K.: Extreme Programming Explained: Embrace Change. 1$^{st}$ ed., Addison-Wesley, Upper Sad- dle River, NJ, 2000.
4. Brooks, F. P.: The Mythical Man- Month – Essays on Software Engineering, Addison-Wesley. 1975.
5. Cirillo, F.: The Pomodoro Technique. XPLabs Technical Report version 1.3. English Version. http://www.tecnicadelpomodoro.it. Published 15 Jun 2007
6. Cramton, C.: The Mutual Knowledge Problem and its Consequences for Dispersed Collaboration. Organization Science 12(3), 2001, 346-371.
7. Espinosa, J.A., Cummings, J.N., Wilson, J.M. and Pearce, B.M.: Team Boundary Issues Across Multiple Global Firms. Journal of Management Information Systems 19(4), 2003, 157-190.
8. Grinter, R.E., Herbsleb, J.D. and Perry, D.E.: The Geography of Coordination: Dealing with Distance in R&D Work. Proc. Int'l ACM SIGGROUP Conf. Supporting Group Work (GROUP '99), ACM Press, New York, 1999, 306-315.
9. Gobbo, F., Vaccari, M.: The Pomodoro Technique for Sustainable Pace in Extreme Programming Teams, In Proceedings of XP2008, Limerick, June 2008
10. Hazzan O, Dubinsky Y. The Software Engineering Timeline : A Time Management Perspective. In: Software-Science, Technology & Engineering, 2007. SwSTE 2007. IEEE International Conference on. Herzlia, Israel; 2007:95-103.
11. Holmstom, H., Fitzgerald, B., Ågerfalk, P.J., and Ó. Conchúir, E.: Agile practices reduce distance in global software development. Information Systems Management Journal Vol 23(3), 2006, 7-18.
12. Humphrey, W. S.: Introduction to the Team Software Process, SEI Series in Software Engineering, Addison-Wesley, 2000.

13. PMI : A Guide to the Project Management Body of Knowledge (PMBOK), 2004.
14. Sarker, S., Sahay, S.: Implications of space and time for distributed work: an interpretive study of US-Norwegian systems development teams. European Journal of Information Systems, 13(1), 2004, 3-20.